\documentclass[12pt]{article}
\usepackage{amsmath}
\usepackage{amssymb}
\allowdisplaybreaks[2]

\renewcommand{\title}[1]{\null\vspace{10mm}\noindent
                         {\Large{\bf #1}}\vspace{10mm}}
\newcommand{\authors}[1]{\noindent{\large #1}\vspace{20mm}}
\newcommand{\address}[1]{{\center{\noindent\small\itshape #1\vspace{0mm}}}}

\def\be{\begin{equation}}
\def\ee{\end{equation}}
\def\bea{\begin{eqnarray}}
\def\eea{\end{eqnarray}}
\def\beb{\begin{eqnarray*}}
\def\eeb{\end{eqnarray*}}
\def\pat{\partial}

\begin{document}
\renewcommand{\baselinestretch}{1.1}
\begin{titlepage}

\begin{center}
\hspace*{\fill}{{\normalsize \begin{tabular}{l}
                              {\sf hep-th/0402229}\\
                              {\sf REF. TUW 04-06}\\
                              \end{tabular} }}

\title{Consistent Construction of Perturbation Theory on 
Noncommutative Spaces}

\authors{Stefan~Denk$^{1a}$, Volkmar~Putz$^{2a}$ and
 Michael~Wohlgenannt$^{3b}$}

\vspace{-15mm}

\address{${}^a$Institut f\"ur Theoretische Physik,
Technische Universit\"at Wien \\
Wiedner Hauptstra\ss e 8--10, A-1040 Wien, Austria\\
\vspace{.3cm}
${}^b$Institut f\"ur Theoretische Physik, Universit\"at Wien\\
Boltzmanngasse 5, A-1090 Wien, Austria}

\footnotetext[1]{denk@hep.itp.tuwien.ac.at}

\footnotetext[2]{putz@hep.itp.tuwien.ac.at}

\footnotetext[3]{miw@ap.univie.ac.at}

\vspace{10mm}

\begin{minipage}{12cm}
{\it Abstract.}
We examine the effect of non-local deformations on the applicability of 
interaction point 
time ordered perturbation theory (IPTOPT) based on the free Hamiltonian of local theories. The usual argument 
for the case of quantum field theory (QFT) on a noncommutative (NC) space (based on the fact that the introduction 
of star products in bilinear terms does not alter the action) is not applicable to IPTOPT due to several discrepancies 
compared to the naive path integral approach when noncommutativity involves time. These discrepancies are explained in 
detail. Besides scalar models, gauge fields are also studied. For both cases, we discuss the free Hamiltonian with 
respect to non-local deformations.

  \vspace*{1cm}
\end{minipage}

\end{center}
\end{titlepage}
\section{Introduction}

It is widely believed that the usual concept of space-time locally modelled as flat Minkowski space breaks down at distances of the magnitude of the Planck scale. One attempt to describe physics at such small scales is to replace the commutative space-time coordinates $x^\mu$ by noncommutative operators $\hat x^\mu$ implying uncertainty relations among the coordinates \cite{Doplicher:1995tu}. The simplest model one can study is characterised by the following commutation relations:
\begin{equation}
  [{\hat x^\mu},{ \hat x^\nu}] = i \theta^{\mu \nu},
\end{equation}
with $\theta^{\mu \nu}$ representing a real, constant, antisymmetric tensor.
We will study this model realised by the so-called (Moyal-Weyl) star product replacing ordinary 
local products of fields within the usual QFT, referred to as NCQFT later on. This star product 
is defined for real-analytic, $L^1$ functions $f$, $g$ as 
\begin{equation}
(f*g)(x) \,\,\equiv\,\,  \left.e^{
        \frac i2 \theta^{\mu\nu}  \partial^\zeta_\mu \partial^\eta_\nu
        }
        f(x+\zeta) g(x+\eta) \right|_{\zeta=\eta=0} \;.
\end{equation}
It is important to note the infinitely many derivatives acting within this product. Concerning QFT, especially the 
time derivatives present for $\theta^{0i} \neq 0$ turn out to be problematic. In that case, a violation of unitarity 
has been observed \cite{Gomis:2000zz} when applying the rules given in \cite{Filk:1996dm}. Unitarity could be 
reestablished in \cite{Bahns:2002vm} using the Yang-Feldmann equation \cite{Yang:1950vi} and in 
\cite{Doplicher:1995tu,Liao:2002xc,Liao:2002pj} by  applying IPTOPT. Below, we
will make clear that this version of time ordering is a consequence of quantum
mechanical basics. IPTOPT was worked out in a more general context
\cite{Denk:2003jj} applicable to a large class of non-local interactions of
scalar particles. The non-locality can be realised by the integral representation of the star product 
\cite{Bozkaya:2002at} as 
\begin{align}
(f \star g)(x) &= \int d^4s \int \frac{d^4 l}{(2\pi)^4} \,
f(x -\tfrac{1}{2} \tilde{l}) \,g(x+s)\, {e}^{ {i} ls}\;,
&  \tilde{l}^\nu := l_\mu \theta^{\mu\nu} \;.
\label{starprod}
\end{align}
This makes the effect of time ordering more transparent. The problems occuring for $\theta^{0i}\neq 0$ can be identified with the 
non-locality in time.
Besides unitarity, the finite UV/IR mixing behaviour \cite{Fischer:2003jh} is a further advantage of IPTOPT. 
At this point, we want to mention that in $\theta$-expanded field theories
\cite{Madore:2000en,Jurco:2001rq,Bichl:2001nf} these difficulties do not appear.
However, they might not be renormalisable \cite{Wulkenhaar:2001sq, Grimstrup:2002af}.

So long, deformed
field theory has been pursued in a somewhat ambivalent way: 
The Moyal product has been used in the interaction part 
of the Lagrangian, the bilinear term remained unchanged \cite{Filk:1996dm} due to the argument 
that in the action one star product can always be omitted:
\begin{equation} 
\int d^4x\,(f \star g)(x)=\int d^4x\,f(x) g(x)\;.
\end{equation}
However, this is not directly applicable to the approach based on IPTOPT.
This is indicated by 
the differences between the Feynman rules given in \cite{Filk:1996dm} and the complicated 
ones based on IPTOPT \cite{Denk:2003jj}. There it was realised that quadratic parts in the 
interaction do have a $\theta$-dependent contribution for $\theta^{0i}\neq 0$. This raises 
the question whether the free Hamiltonian in the framework of IPTOPT gets deformed or not 
when introducing the star product (\ref{starprod}) and i.g.
non-local deformations. We will adress this problem for scalar as well as gauge 
field models. We will also construct IPTOPT beginning with the Schr\"odinger equation to make 
the definition of time ordering clear.  The discrepancies between IPTOPT and a naive path 
integral approach giving the simple Feynman rules \cite{Filk:1996dm} will be explained.

In Section $5$, we will discuss the canonical deformation of gauge field theory and 
BRST-Symmetry also in the Hamiltonian approach.

      
\section{Quantum Mechanics}
In this section, quantum mechanical basics needed later are repeated. After the introduction of various types of time dependence of operators and states time ordered perturbation theory is discussed.
\subsection{Schr\"o{}dinger Picture}

We start with the Schr\"o{}dinger equation (with $\hbar =1 $),
\begin{equation}
i\frac\partial{\partial t}|Z_S(t)\rangle = H_S|Z_S(t)\rangle, \label{schroe}
\end{equation}
where $H_S$ 
is the Hamilton operator and $|Z_S(t)\rangle$ is a time 
dependent Schr\"odinger state.
As long as $H_S$ is time independent, we have the simple solution
\begin{equation}
|Z_S(t)\rangle = e^{-iH_S(t-t_0)}|Z_S(t_0)\rangle,
\end{equation}
with some initial state $|Z_S(t_0)\rangle$.

Now the question is: What is the particular feature of a specific model
described by eq.~(\ref{schroe})? The answer is simple: Different models are
distinguished by different Hamiltonians. In particular, the states are 
defined as solutions of eq.~(\ref{schroe}) with a particular Hamiltonian
$H_S$.

In the Schr\"o{}dinger picture (which is denoted by the index $S$) we
have the notion of a time independent
Hamiltonian which generates time dependent states. Physics is described
via matrix elements of operators with these states. Those operators are 
assumed time independent and (if we are lucky) known, so the interesting
thing is to get the correct states. 

\subsection{Heisenberg Picture}

But if we have a closer look at the matrix elements of some time independent
operator $A_S$,
\begin{equation}
\langle A\rangle = \langle Z_S(t)| A_S | Z_S(t)\rangle = 
\langle Z_S(t_0)|e^{+iH_S(t-t_0)} A_S e^{-iH_S(t-t_0)} | Z_S(t_0)\rangle, 
\end{equation}
we could also argue that we have time independent states $|Z_H\rangle:=
|Z_S(t_0)\rangle$
and time dependent $Heisenberg$ operators $A_H$ (let $t_0 = 0$),
\begin{equation}\label{heisop}
A_H(t) :=e^{+iH_S t} A_S e^{-iH_S t}.
\end{equation}
This is the Heisenberg picture. Instead of fixing the operators and 
searching for time dependent states, we keep the states fixed and put our 
interest in time evoluting operators.
 
Instead of the Schr\"o{}dinger equation (\ref{schroe}) for the states, we
now have the Heisenberg equation for the Heisenberg operators, obtained from 
differentiating (\ref{heisop}) with respect to the time 
(note that $A_S$ does not explicitly depend on time),
\begin{equation}
-i\frac\partial{\partial t}A_H(t) = [H_H, A_H(t)].
\end{equation}
Here, $H_H = H_S$ is still time independent. 
The Heisenberg equation looks indeed very similar to the
Schr\"o{}dinger equation.

\subsection{Dirac Picture}

Somehow in between is the Dirac picture, where states and operators
have a time evolution. The $free$ part $H_{0S}$
of the Hamiltonian $H_S=H_{0S}+H_{IS}$ is used to describe 
the time evolution of the operators,
whereas the $interaction\ part$ $H_{IS}$ will describe 
the time evolution of the states.
The states in the interaction picture
are defined as
\begin{eqnarray}\label{dstate}
|Z_D(t)\rangle :=e^{iH_{0S}t}|Z_S(t)\rangle\ .
\end{eqnarray}
From $\langle Z_S | A_S | Z_S\rangle = \langle Z_D | A_D | Z_D\rangle$ we 
conclude
\begin{equation}
A_D(t) := e^{+iH_{0S}t}A_Se^{-iH_{0S}t}\ . 
\label{doper}
\end{equation}
With 
\begin{eqnarray}
&H_{0D} = H_{0S},\quad H_{ID}(t) :=  e^{+iH_{0S} t} H_{IS} e^{-iH_{0S} t},
\nonumber\\
&[H_{0S}, \exp(\pm i H_{0S} t)] = 0 ,\\
&H_{ID}(t=0)=H_{IS}\nonumber
\end{eqnarray}
we find the two evolution equations (from (\ref{schroe}), 
(\ref{dstate}),
respectively)
\begin{eqnarray} \label{evol}
&&-i\frac\partial{\partial t}A_D(t) = [H_{0D}, A_D(t)],\nonumber\\
&&i\frac\partial{\partial t} |Z_D(t)\rangle = H_{ID}(t)|Z_D(t)\rangle.
\end{eqnarray}
We see that
the time evolution of the operators is defined by the $free$ Hamiltonian,
so that $A_D(t)$ is simply a solution of the free theory. The time 
evolution of the states, on the other hand, depends only on the
$interaction$ Hamiltonian. Note: Since the free Hamiltonians
in the Schr\"odinger and Dirac picture are the same, we define 
$H_{0S}  = H_{0D} \equiv H_0$ from now on.

\subsection{IPTOPT}
Given some (non-local) interaction Hamiltonian $H_{ID}(t)$ in the Dirac picture,
the evolution equation (\ref{evol}) for the Dirac states 
can be solved by introducing the time evolution operator $U(t,t_0)$ as
\begin{equation}
U(t,t_0)|Z_D(t_0)\rangle \equiv |Z_D(t)\rangle \;.
\end{equation}
The well known solution of the resulting differential equation with the boundary condition $U(t_0,t_0)=1$ is
\begin{eqnarray} \label{nobelprize}
U(t,t_0)=\sum_{n=0}^\infty \frac{(-i)^n}{n!}
   \int_{t_0}^t dt_1\ldots dt_n \,T\{H_{ID}(t_1)\ldots H_{ID}(t_n)\}
   \nonumber \\\label{Utt0}\equiv Te^{-i\int_{t_0}^t H_{ID}(t')dt'}\;.
\end{eqnarray}
$T$ is the time ordering operator and it should be pointed out here that it rearranges 
the whole operators $H_{ID}(t)$ according to their time argument $t$ and it does not act 
on parts of  $H_{ID}(t)$. We call this \emph{interaction point time ordering} (IPTO) to 
distinguish it from other possible time orderings of objects.
Since we are considering non-local deformations of the classical theory, the concept of
interaction at a point is no longer valid. By {\it interaction point} we mean the time 
argument of the interaction Hamiltonian in Eq.~(\ref{nobelprize}). The interaction 
Hamiltonian may of course be a non-local expression and contain products of fields
at different times, cf. (\ref{starprod})

In order to describe
scattering processes, we need the  S-operator
defined by 
\begin{equation}
S \equiv U(\infty,-\infty)=Te^{-i\int dt \,H_{ID} (t)}\;.
\end{equation}
Again, $T$ acts on the time argument $t$ of 
$H_{ID}(t)$.
Only for simple theories such as scalar $\phi^4$ theory, one can write 
$ H_{ID} (t)= - \int d^3x\,{\mathcal L}_{ID} (t,\vec x) $. But note that IPTOPT principally requires the interaction Hamiltonian and not the Lagrangian.

S-matrix elements are thus given by
\begin{equation}
S_{fi} :=\langle f|S|i\rangle,
\end{equation}
where $|i\rangle$ and $\langle f|$ represent the incoming and outgoing states,
respectively.


\section{Quantum Field Theory}
The operators we have dealt with in the last section can usually be expressed in terms of field operators $\phi$ and their canonical conjugates $\pi$. 
Their operator character is manifested by the equal time commutation relations
\begin{eqnarray}
[\pi(t,\vec x),\pi(t,\vec x')] &=& [\phi(t,\vec x),\phi(t,\vec x')]=0,
\nonumber\\[0cm]
 [\pi(t,\vec x),\phi(t,\vec x')]&=&
 -i\delta^3(\vec x-\vec x').\label{eq:quanti0}
\end{eqnarray}
For the application of IPTOPT the free field operators, which are given in the Dirac picure, are especially important. 
Their dynamics can be characterised by the Lagrange or Hamilton formalism, as we will briefly discuss in this section.
The main investigation of this section is the effect of deformation on the free Hamiltonian $H_0\rightarrow H_0^*$.
Star products in the action can be left out for quadratic terms, which is often
given as the reason that the free theory on noncommutative spaces is the same as
for the commutative one. In the Hamiltonian, however, the star product does not automatically (only for $\theta^{0i}=0$) 
drop out of quadratic terms in the interaction. This is the reason  why we adress the deformation of the free 
Hamiltonian from another point of view.

\subsection{Commutative Space}

The free scalar field is described by the equation of motion
\begin{equation}
(\square + m^2)\phi = 0.
\end{equation}
This equation of motion can be obtained by a field variation of 
the action,
\begin{eqnarray}
&&W=\int dt L_0 = \int d^4x {\mathcal L}_0,\quad {\mathcal L}_0=
\frac12(\partial^\mu\phi)(\partial_\mu\phi)-\frac12 m^2\phi^2,\nonumber\\
&&\delta W = 0 \Rightarrow (\square +m^2)\phi = 0.
\end{eqnarray}
Another possibility is the description via the Hamiltonian 
$H=\int d^3x {\mathcal H}$,
\begin{eqnarray}
\pi:= \frac{\partial {\mathcal L}}{\partial \dot\phi} = \dot\phi,\quad
{\mathcal H} := \dot\phi\pi - {\mathcal L}.
\end{eqnarray}
An explicit calculation yields
\begin{equation}
H_0 =\int d^3x \frac12 (\dot\phi^2 + (\vec\partial\phi)^2 +m^2\phi^2) \geq 0.
\label{ham}
\end{equation}
$H_0$ is interpreted as the total energy of the system.
Energy conservation $\frac d {dt}H_0 = 
\frac d {dt}\int d^3x \, {\mathcal H}_0 = 0$ is obtained by use of partial 
integration and the
equation of motion. 

Note: $\phi$ and $\pi$ correspond to $x$ (the current elongation) and 
$p$ (the momentum) of the harmonic oscillator, 
whereas the space coordinates $\vec x$ could be thought of as 'labels' of
the infinitely many harmonic oscillators hanging around in space. Only time
is always time.

Since $\phi$ satisfies the free field equation, we can write it as \\
$\phi(x) = \phi^+(x) +\phi^-(x)$, where
\begin{eqnarray}\label{phi+-}
&&\phi^-(x) = \frac1{(2\pi)^{3/2}}\int\frac{d^3k}{\sqrt{2\omega_k}}
\;a^-(\vec k)\;e^{-ix_\mu k^{+\mu}}, \nonumber\\
&&\phi^+(x) = \frac1{(2\pi)^{3/2}}\int\frac{d^3k}{\sqrt{2\omega_k}}
\;a^+(\vec k)\;e^{+ix_\mu k^{+\mu}}.
\end{eqnarray}
Here we have $k^+ = (\omega_k, \vec k),\ \omega_k = \sqrt{\vec k^2+m^2}$.
The combination with eq.~(\ref{eq:quanti0}) yields
\begin{equation}\label{quanti}
[a^-(\vec k),a^+(\vec k')]=
\delta^3(\vec k - \vec k'). 
\end{equation}
Finally the Hamiltonian can be  rewritten as 
\begin{eqnarray} \label{H0}
H_0=\int d^3k\,\omega_k\,\frac12(a^+(\vec k)a^-(\vec k) +
a^-(\vec k)a^+(\vec k)).
\end{eqnarray}

%
%

\subsection{Non-local Deformation}

In this subsection, we want to study the effect of general non-local deformations on the 
free scalar Hamiltonian. At the end, we will examine some examples, such as canonical 
deformation.
Let us start with the usual Hamiltonian
\begin{align}\label{hamilton}
&H_0  
=\int d^3x \frac12(\dot\phi^2 + (\vec\partial \phi)^2 + m^2\phi^2).
\end{align}
Now, let us introduce a non-local deformation of the above pointwise product:
\be
f(x)g(x) \to \int d\underline{\mu}\, w(\underline{\mu}) \, f(x+h_1(\underline{\mu}))\,
g(x+h_2(\underline{\mu})).
\ee
For $H_0$ one thus gets
\bea
H_0(t) \to H_0^*(t) & = & \frac12 \int d^3 x \int d\underline{\mu}\, w(\underline{\mu}) \times
\\
&& 
\hspace{-2.5cm}
\left( \pat^\nu \phi(x+h_1(\underline{\mu}))\pat^\nu \phi(x+h_2(\underline{\mu})) + m^2
\phi(x+h_1(\underline{\mu}))\phi(x+h_2(\underline{\mu})) \right).
\nonumber
\eea
Still, $\phi(x)$ 
shall denote the
free field operator obeying the usual free field equation
with physical mass $m$ given by
\be
\pat^\mu \pat_\mu \phi - m^2 \phi =0.
\ee
Therefore, we can apply Fourier transformation and interpret the coefficients as creation and 
annihilation operators $a^\dagger (\mathbf k)$ and $a (\mathbf k)$, respectively. Straight 
forward calculation yields
\bea
H_0^*(t) & = & \frac12 \int d^3 k \int d\underline{\mu}\, w(\underline{\mu}) \times
\\
\nonumber
&& \left(
a(\mathbf k) a^\dagger (\mathbf k) e^{ik^+(h_1(\underline{\mu})-h_2(\underline{\mu}))}+
a^\dagger(\mathbf k) a(\mathbf k) e^{-ik^+(h_1(\underline{\mu})-h_2(\underline{\mu}))}
\right).
\eea
The coefficients of the terms proportional to 
$a^\dagger(\mathbf p)a^\dagger(\mathbf k)$
and
$a(\mathbf p)a(\mathbf k)$
vanish:
$$
(p^{+\mu}k^{+\mu}+m^2)\delta^3(\mathbf{k+p})=0,
$$
whereas the coefficient of $a(\mathbf p)a^\dagger(\mathbf k)$ and $a^\dagger(\mathbf p)a(\mathbf k)$
is proportional to
$$
(-p^{+\mu}k^{+\mu}+m^2)\delta^3(\mathbf{k+p})=2 \omega_p^2 \delta^3(\mathbf{p+k}).
$$
Usually, the Hamiltonian is normal ordered. For $H_0^*$ we obtain
\be
:H_0^*(t): = \int d^3k\, \omega_k \, a^\dagger(\mathbf k) a(\mathbf k) \xi(\mathbf k),
\ee
where $\xi$ is given by
\be
\xi(\mathbf k) = \int d\underline{\mu}\, w(\underline{\mu})\, \cos (k^+(h_1^(\underline{\mu})-h_2(\underline{\mu}))).
\ee
If $\xi(\mathbf k)$ is constant, $H_0^*$ and $H_0$ only differ by an overall normalisation
constant. Therefore, the free Hamiltonian would be unaltered. But: What deformations
do that job?

A trivial solution to this question is the choice $h_1(\underline{\mu})=h_2(\underline{\mu})$ and the 
requirement $\int d(\underline{\mu}) \,w(\underline{\mu})<\infty$. Actually this deformation 
is still local.

The next example is canonical deformation, discussed in the Introduction. We have
$\underline{\mu}=\{l,s\}$, $w(\underline{\mu})=\exp (isl)/(2\pi)^4$, $h_1(\underline{\mu})=-\frac12 \tilde l$
and $h_2(\underline{\mu})=s$. Thus, we obtain for $\xi$
\be
\xi(\mathbf k) = e^{-ik^+\tilde k^+/2} = 1,
\ee
and the free Hamiltonian is unaltered by the deformation \cite{Doplicher:1995tu}. The use of the perturbation 
theory worked out in \cite{Liao:2002xc,Denk:2003jj,Bozkaya:2002at}
is justified.

As a third example, we consider the approach to UV-finite theories considered in \cite{Bahns:2003vb}.
Comparison with our definitions yields
\beb
& \underline{\mu}=\{a_1, a_2\}, \, h_1(\underline{\mu})=\zeta a_1, \, h_2(\underline{\mu})=\zeta a_2,\\
& w(\underline{\mu}) = 2c_2 e^{\frac12 (a_1^{\nu 2}+a_2^{\nu 2})} \delta^4(a_1+a_2).
\eeb
Therefore, we have
\be
\xi(\mathbf k) = 2 c_2\pi^2 e^{-\zeta^2k_\nu^{+2}},
\ee
where $\zeta$ has dimension of length in order to keep the parameters $a_i$
dimensionless and to provide control over the non-locality, and $c_2$ represents
a normalization constant.
With a suitable choice for $c_2$ we get the following Hamiltonian:
\be
:H_0^*(t): = \int d^3k\, \omega_k' \, a^\dagger(\mathbf k) a(\mathbf k) \xi(\mathbf k),
\ee
where
\be
\omega_k' = \omega_k e^{-\zeta^2 k^{+2}_\nu}. 
\ee
In principle, this Hamiltonian can be interpreted as the new free particle energy.

The simplified UV-finite QFT introduced in \cite{Denk:2004pk} gives a similar result. 
We have $\underline{\mu} = \{b_1,
b_2 \}$, $h_1(\underline{\mu})=M \, b_1$ and $h_2(\underline{\mu})= M \, b_2$. 
Hence, $\xi$ is given by
\be
\xi(\mathbf k) = e^{-\frac12 k^{+T}\kappa\, k}.
\ee

In the last two examples the deformed free Hamiltonian does not equal the undeformed one. 
Also, the interpretation of $ \omega_k'$ as the energy of a physical state with momentum
$ \mathbf k$ is troublesome, since the energy goes to zero for large $\mathbf k$.
Therefore, the deformation has only been introduced in the interaction terms in 
\cite{Bahns:2003vb} and \cite{Denk:2004pk}, respectively.


\section{A Conceptual Note}

In this section, we want to discuss discrepancies between IPTOPT and the naive path integral approach (NPIA). 
By IPTOPT we mean calculations according to eq.~(\ref{Utt0}) with 
\begin{equation}\label{eq:HID}
H_{ID}(x^0) \equiv \frac{\lambda}{k!}\int d^3x\,\,(\phi*)^k(x),
\end{equation}
and $\phi$ denoting the field operators in the Dirac picture.
The resulting Feynman-rules are given in 
\cite{Denk:2003jj}.
By NPIA we mean that one calculates $n$-point functions according to the path integral
\begin{equation}
   \int {\mathcal D\phi}\;\phi(x_1)\ldots\phi(x_n)\;e^{iI[\phi]}\;,
\end{equation}
with $I$ denoting the corresponding action including $i\epsilon$ terms:
\begin{equation}
I[\phi] =\int d^4x\,\,
   \left[{\mathcal L}_0+\frac{\lambda}{k!}(\phi*)^k(x)+i \epsilon\right].
\end{equation}
The corresponding Feynman rules are the same as for the local theory but with phase factors to be included for vertices \cite{Filk:1996dm}. Meanwhile, it is clear that these two approaches differ when noncommutativity involves time.
The most striking problem is the unitarity violation 
\cite{ Gomis:2000zz} when applying these naive Feynman rules, which can be cured by a strict application of IPTOPT 
\cite{Bahns:2002vm,Liao:2002pj}. Another mismatch was realised in 
\cite{Denk:2003jj}, where it turned out that the star product of NCQFT does not drop out of some quadratic terms which might be considered as counter 
terms for example. But in the Lagrangian (NPIA), star products become redundant in  any bilinear term. 

For the  local version of the model we are studying, both, the NPIA and IPTOPT give the same results. But for non-local theories as NCQFT, these approaches are not equivalent. To see the crucial points about these mismatches, we sketch how to pass from IPTOPT to the NPIA \cite{Weinberg}. One starts with the Hamiltonian $H_S\equiv H_D(0)$ written as a functional of canonical field operators $\phi_D(t,\vec x), \pi_D(t,\vec x)$ in the Dirac picture
\begin{equation}\label{eq:Hfunc}
H_D(t)\equiv H[\phi_D,\pi_D;t]\;.
\end{equation}
To be specific, our version of IPTOPT and NCQFT gives 
\begin{equation}
H_D(t)= H_0+H_{ID}(t)
\end{equation}
combined with eq.~(\ref{eq:HID}).
The special notation of the functional (\ref{eq:Hfunc}) is due to the non-localities, 
especially the ones in time: it is not a functional depending just on field operators 
given at fixed time $t$, but all possible times are involved (see also eq.~(\ref{phi4}) 
below).
Furthermore, one assumes two complete basis sets $|q;t\rangle$ and $|p;t\rangle$ 
for each time $t$ being eigenstates of  the field operators in the Heisenberg picture 
$\phi_H(t,\vec x)$ and $\pi_H(t,\vec x)$, respectively. The goal is to evaluate 
scalar products $\langle q';t'|q;t\rangle$ between basis vectors given at different 
times $t<t'$. Then, one sandwiches sums over complete basis sets belonging to 
intermediate times $t_i$ with $ t<t_1\ldots <t_N<t'$. So far, everything might 
also work for non-local field theories. Next, one has to evaluate matrix elements like
\begin{equation}
\langle q';t+dt|q;t\rangle = \langle q';t|e^{-iH_Sdt}|q;t\rangle.
\end{equation}
At this point, $H_S$ is usually rewritten
\begin{equation}
H_S= e^{iH_St}e^{-iH_St}H[\phi_D,\pi_D;0]
 =e^{iH_St}H[\phi_D,\pi_D;0]e^{-iH_St}\;.
\end{equation}
In a local theory, the functional just depends on field operators $\phi_D(0,\vec x)$ evaluated at time $t=0$ which can be simply replaced using $\phi_D(0,\vec x)=\phi_H(0,\vec x)$.
As a matter of course, one thus rewrites this by sandwiching unit operators $\exp(iH_St)\exp(-iH_St)$ as
\begin{equation}\label{eq:HD2HH}
e^{iH_St}H[\phi_D,\pi_D;0]e^{-iH_St}=e^{iH_St}H[\phi_H,\pi_H;0]e^{-iH_St}
=H[\phi_H,\pi_H;t]
\;.
\end{equation}
However, these steps are problematic for theories which are non-local in time.
To see this, we consider an operator $O_D$ defined as a Moyal product of two operators $A_D$, $B_D$ in the Dirac picture:
\begin{equation}
O_D(x)\equiv(A_D \star B_D)(x) = \int d^4s \int \frac{d^4 l}{(2\pi)^4} \,{e}^{ {i} ls}\,
A_D(x -\tfrac{1}{2} \tilde{l}) \,B_D(x+s) \;.
\label{starprod1}
\end{equation}
The subscript $D$ at $O_D$ indicating that $O_D$ has the time dependence of an operator in the Dirac picture is justified since
$$
O_D(t,\vec x)=e^{iH_0 t}O_D(0,\vec x)e^{-iH_0 t}
$$
holds. The transition from the Dirac  to the Heisenberg picture is now done as usual \cite{Weinberg}:
\begin{equation}
  O_H(t,\vec x)=e^{iH_S t}O_D(0,\vec x)e^{-iH_S t}\,.
\end{equation}
Substituting eq.~(\ref{starprod1}), we get
\begin{eqnarray}\nonumber
  O_H(t,\vec x)&=&\int d^4s \int \frac{d^4 l}{(2\pi)^4} \,{e}^{ {i} ls}\,
  e^{iH_S t}A_D(x_0-\frac12 \tilde l)B_D(x_0+s)e^{-iH_S t}
\\
&=&
\int d^4s \int \frac{d^4 l}{(2\pi)^4} \,{e}^{ {i} ls}\,
W(t,t-{\tilde l}^0/2) 
A_H(x-{\tilde l}/2)\times\label{eq:nlprod}\\&&
W(t-{\tilde l}^0/2,t+s^0)
B_H(x+s)
W(t+s^0,t)
\,.\nonumber
\end{eqnarray}
where $x=(t,\vec x)$, $x_0=(0,\vec x)$, and
\begin{equation}
  W(t,t_0)\equiv e^{iH_S t}e^{-iH_0 (t-t_0)}e^{-iH_S t_0}\,.
\end{equation}
$W(t,t_0)$ is unitary and $W(t,t)=1$, but in general $W(t,t_0)$ is not the unit operator.
In order to 
stay consistent one thus has to redefine the 
noncommutative product with respect to Heisenberg fields correspondingly. 
For the Hamiltonian needed for path integrals, one could proceed with
\begin{equation}
e^{iH_S t}H[\phi_D,\pi_D;0]e^{-iH_S t}
=:
H'[\phi_H,\pi_H;t].
\end{equation}
When non-locality involves time clearly $H'\neq H$, and it can be expected that dealing with $H'$ is pretty hard.
We assume that $H'\neq H$ is the cause for the discrepancies between IPTOPT and the NPIA for systems described by Hamiltonians which are non-local in time.
Further problems are expected when integrating out the conjugate momenta to pass from the Hamiltonian to the Lagrangian formulation, even if we accept $H$ instead of $H'$
\footnote{Clearly, this would not be equivalent to our IPTOPT approach. 
It would mean that we started with $H_S=H[\phi_H,\pi_H;t=0]$. }. This might be due 
to the fact that non-locality in time means that $H$ depends on infinitely many 
time derivatives implying complicated equations of motion \cite{Ostrogradski}.
Furthermore, the equivalence of using  the Lagrangian interaction or the 
Hamiltonian is not justified anymore by path integrals. We believe that it is 
important to check this explicitly. Besides derivative couplings and 
counterterms, NCQED might also be affected due to the complicated quantisation 
procedure which involves derivatives through the constraints \cite{Dirac}.

Before dealing with gauge field models in the following section, we want to
illustrate the differences of the NPIA and IPTOPT. The perturbation expansion of
the NPIA is obtained by expanding the integrand in terms of the interaction
leaving only the bilinear parts in the exponential. The resulting path integrals
can be carried out and one gets Feynman rules which associate the usual
propagators $\Delta_F(x,y)$ (inverse of the bilinear parts) of the local theory with 
lines, and vertices contain four-momentum dependent phase factors. A 
time ordering interpretation of the resulting rules can be obtained by writing 
$\Delta_F(x,y)$ in terms of time ordered products of the free annihilation and 
creation fields, $\phi^-$ and $\phi^+$, respectively:
\begin{equation}
\Delta_F(x,y)=\tau(x^0-y^0)[\phi^-(x),\phi^+(y)]
             +\tau(y^0-x^0)[\phi^-(y),\phi^+(x)]\;.
\end{equation}
Here, $\tau(t)$ is the time ordering step function $\tau(t)=1$ for $t> 0$ and
$\tau(t)=0$ for $t < 0$.
This indicates that the NPIA can be interpreted in the sense of a  \emph{total} time ordering acting with respect to the time argument of each field. 
 On the other hand, the time ordering operator of IPTOPT just rearranges whole 
 interactions. Now, let us consider $\phi^4$-theory. We have
\begin{eqnarray}\label{phi4}
&&(\phi\star\phi\star\phi\star\phi)(z) = 
\int \prod_{i=1}^3 \Big(d^4s_i\frac{d^4 l_i}{(2\pi)^4}e^{il_i s_i}\Big)
\\&&
\quad \times\phi(z-\frac12\tilde l_1)\phi(z+s_1-\frac12\tilde l_2)
\phi(z+s_1+s_2-\frac12\tilde l_3)\phi(z+s_1+s_2+s_3),\nonumber
\end{eqnarray}
which clearly expresses the non-locality. The two-point function at first order 
in $\lambda$ can then be written a bit sloppy as
\begin{equation}
G(x,y) = \frac{g}{4!} \int d^4z \,\Big\langle0\Big| 
T \big( \phi(x) \phi(y) \big(\phi
\star \phi \star  \phi \star \phi\big)(z)\big) \Big|0\Big\rangle_{(0)} \;.
\label{Gxy}
\end{equation}
Let us
discuss the total and the IPTOPT time ordering  for one particular geometrical situation with respect to
(\ref{Gxy}):
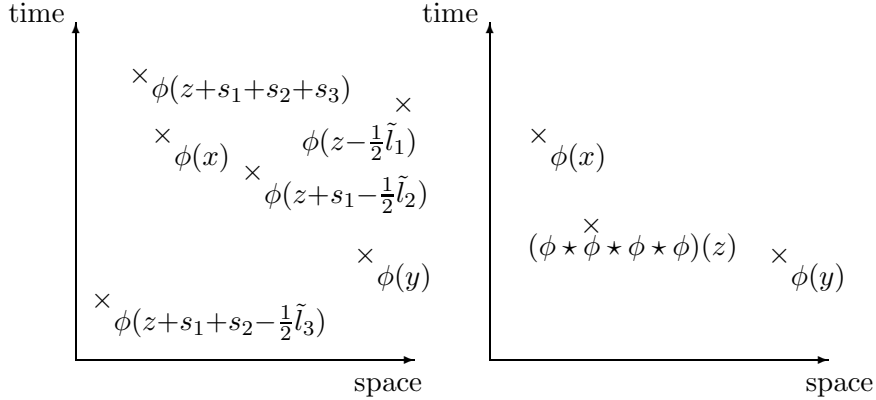
\begin{align}\nonumber
\parbox{100mm}{\begin{picture}(100,50)
\put(3,3){\vector(1,0){45}}
\put(3,3){\vector(0,1){45}}
\put(-6,48){\mbox{\small time}}
\put(40,-1){\mbox{\small space}}
\put(5,10){\mbox{\small$\times$}}
\put(8,7){\mbox{\small$\phi(z{+}s_1{+}s_2{-}\tfrac{1}{2} \tilde{l}_3)$}}
\put(45,36){\mbox{\small$\times$}}
\put(33,31){\mbox{\small$\phi(z{-}\tfrac{1}{2} \tilde{l}_1)$}}
\put(25,27){\mbox{\small$\times$}}
\put(28,24){\mbox{\small$\phi(z{+}s_1{-}\tfrac{1}{2} \tilde{l}_2)$}}
\put(10,40){\mbox{\small$\times$}}
\put(13,38){\mbox{\small$\phi(z{+}s_1{+}s_2{+}s_3)$}}
\put(40,16){\mbox{\small$\times$}}
\put(43,13){\mbox{\small$\phi(y)$}}
\put(13,32){\mbox{\small$\times$}}
\put(16,29){\mbox{\small$\phi(x)$}}
\put(58,3){\vector(1,0){45}}
\put(58,3){\vector(0,1){45}}
\put(49,48){\mbox{\small time}}
\put(100,-1){\mbox{\small space}}
\put(70,20){\mbox{\small$\times$}}
\put(63,17){\mbox{\small$(\phi\star \phi\star\phi\star\phi)(z)$}}
\put(95,16){\mbox{\small$\times$}}
\put(98,13){\mbox{\small$\phi(y)$}}
\put(63,32){\mbox{\small$\times$}}
\put(66,29){\mbox{\small$\phi(x)$}}
  \end{picture}}
\end{align}
The arrangement of fields for the left figure 
corresponds to the following non-vanishing
contribution to the total time ordering of (\ref{Gxy}):
\begin{align}
&G(x,y)=\nonumber\\ &= 
\int d^4z  \int \prod_{i=1}^3 \Big( d^4 s_i \frac{d^4 l_i}{(2\pi)^4} 
\; {e}^{ {i}l_i s_i}\Big)
\tau(s_1^0{+}s_2^0{+}s_3^0{+}\tfrac{1}{2}\tilde{l}_1^0)
\tau(z^0{-}\tfrac{1}{2}\tilde{l}_1^0{-}x^0)
\nonumber
\\ 
&
\times 
\tau(x^0{-}z^0{-}s_1^0{+}\tfrac{1}{2}\tilde{l}_2^0)
\tau(z^0{+}s_1^0{-}\tfrac{1}{2}\tilde{l}_2^0{-}y^0)
\tau(y^0{-}z^0{-}s_1^0{-}s_2^0{+}\tfrac{1}{2}\tilde{l}_3^0)
\label{GTxy}
\\ 
&
\times\Big\langle0\Big| 
\phi (z{+} s_1{+}s_2{+}s_3) 
\phi(z{-}\tfrac{1}{2} \tilde{l}_1) \phi(x)
\phi(z{+}s_1{-}\tfrac{1}{2} \tilde{l}_2) 
\phi(y) 
\phi(z{+}s_1{+}s_2{-}\tfrac{1}{2} \tilde{l}_3) 
\Big|0\Big\rangle_{(0)}.
\nonumber
\end{align}
We find that there are $6!=720$ different 
contributions to (\ref{Gxy}) when interpreting the time ordering in the 
Gell-Mann--Low formula as a total time ordering of all field arguments, as 
one would expect. This kind of time ordering guarantees that only causal
processes contribute to the $S$-matrix. 

In contrast to this {\it total} time ordering, we now 
have $interaction\ point$ time ordering (right figure), which is
defined with respect to the \emph{interaction point}:
\begin{align}
&G'(x,y) = 
\int d^4z  \int \prod_{i=1}^3 \Big( d^4 s_i \frac{d^4 l_i}{(2\pi)^4} 
\; {e}^{ {i}l_i s_i}\Big)
\tau(x^0{-}z^0) \tau(z^0{-}y^0) 
\label{GTPxy}
\\ 
& 
\times\!\Big\langle0\Big| 
\phi(x) \phi(z{-}\tfrac{1}{2} \tilde{l}_1) 
\phi(z{+}s_1{-}\tfrac{1}{2} \tilde{l}_2) 
\phi(z{+}s_1{+}s_2{-}\tfrac{1}{2} \tilde{l}_3) 
\phi (z{+} s_1{+}s_2{+}s_3) \phi(y)
\Big|0\Big\rangle_{(0)}.
\nonumber
\end{align}
There are now only $3!=6$ different contributions of this type. For most 
contributions some of the fields are now  at the ``wrong''
place with respect to the total time order. Thus the noncommutative version 
(\ref{GTPxy}) of the Gell-Mann--Low formula violates causality but preserves
unitarity (as we want to stress once more). After all, 
contributions to the Dyson series are precisely ordered only with respect to 
the time stamp of the interaction Hamiltonians.

\section{Gauge Field Theory}

In this section, we will compute the noncommutative Hamiltonian for pure gauge 
theory, with and without ghosts. For simplicity, we restrict ourselves to the 
case of $U(1)$ gauge theory. We also do not employ Seiberg-Witten maps, but only
replace the pointwise product of fields with the $*$-product~(\ref{starprod}).

\subsection{Gauge Fixed Lagrangian}

The free part of the pure $U(1)$ gauge field Lagrangian on commutative space
reads
\begin{equation}
{\mathcal L}_0 =  -\frac 14 f_{\mu\nu}f^{\mu\nu} - \frac 1{2\alpha}
(\partial^\mu A_\mu)(\partial^\nu A_\nu),
\end{equation}
where we have defined
\begin{equation}
f_{\mu\nu} :=\partial_\mu A_\nu-\partial_\nu A_\mu.
\end{equation}
The free field equation reads
\begin{equation}
\square A_\mu -(1-\frac1\alpha)\partial_\mu(\partial^\nu A_\nu)=0.
\label{eomgf}
\end{equation}
For the free field momenta we find
\begin{eqnarray}
\Pi^i = f^{i0} = +f_{0i},\qquad \Pi^0 = -\frac1\alpha(\partial^\mu A_\mu).
\end{eqnarray}
Thus we define the \emph{noncommutative} Hamiltonian
\begin{eqnarray}
H_0^*&=& \int d^3 x \, \mathcal H_0^* \equiv
\int d^3x(\frac12\{\dot A_\mu, \Pi^\mu\}_\star -
{\mathcal L}_0^\star)\nonumber\\
& =& \int d^3x \frac12\{\partial_0 A_i, f_{0i}\}_\star
+\hspace{-1.7cm}\underbrace{\frac12\partial^\mu A^\nu\star f_{\mu\nu}}_{
-\frac14\{\partial_0A_i,f_{0i}\}_\star-\frac14\{\partial_i A_0,f_{i0}\}_\star
+\frac12\partial_i A_j\star f_{ij}}\nonumber\\&&\qquad
-\frac1{2\alpha}\{\partial_0 A_0,\partial^\mu A_\mu\}_\star+ \frac1{2\alpha}
(\partial^\mu A_\mu)\star(\partial^\nu A_\nu)\nonumber\\
&=&\int d^3x\frac14\{(\partial_0 A_i+\partial_i A_0),f_{0i}\}_\star
+\hspace{-0.8cm}\underbrace{\frac12\partial_i A_j\star f_{ij}}_{
\frac12(\partial_i A_j\star\partial_i A_j- \partial_i A_i
\star\partial_j A_j)} \nonumber\\&&\quad
-\frac1{2\alpha}(\partial_0 A_0)
\star(\partial_0 A_0) +\frac1{2\alpha}(\partial_i A_i)\star(\partial_j A_j) 
+\dot A_0\star\dot A_0
-\dot A_0\star\dot A_0
\nonumber\\
&=&\frac12\int d^3x \Big(\dot A_i\star \dot A_i -\dot A_0\star \dot A_0
+\partial_j A_i \star\partial_j A_i-\partial_j A_0 \star\partial_j A_0
\nonumber\\&&\qquad
+(\frac{\alpha-1}{\alpha})(\dot A_0\star \dot A_0
-(\partial_i A_i)\star(\partial_j A_j))\Big)\label{Hamiltongf}.
\end{eqnarray}
We check explicitly the time independence of the Hamiltonian,
\begin{eqnarray}
\dot H_0^* &=& \frac 12\int d^3 x \frac12\{\dot A_i,\ddot A_i-
\vec\partial^2 A_i+(\frac{\alpha-1}{\alpha})(-\partial_i\partial^\mu A_\mu+
\underbrace{\partial_i \partial^0 A_0})\}_\star\nonumber\\&&
-\frac12\{\dot A_0,\ddot A_0-\vec\partial^2 A_0 -
(\frac{\alpha-1}{\alpha})\partial_0\partial^0 A_0\}_\star,
\end{eqnarray}
where $\vec\partial^2 = \partial_j \partial_j$.
Note: The use of the Moyal-anticommutators is crucial!
The first line (without the underbraced term) is zero due to the equation
of motion for $A_i$. After partial integration of the underbraced term 
with respect to $\partial_i = -\partial^i$ it combines
with the second line to the equation of motion for $A_0$. Thus we see that 
$\frac d{dt} H_0^* =0$.

For quantisation we rewrite $H_0^*$ in a convenient form
\begin{eqnarray}
H_0^*=\frac12 \int d^3x\Big(-\partial_\mu A^\nu\star\partial_\mu A_\nu
+(\frac{\alpha-1}{2\alpha})\{\partial^\mu A_\mu, \partial_\nu A_\nu\}_\star
\Big).
\end{eqnarray}
We make the ansatz
\begin{equation}
A_\mu(x)= 
\frac1{(2\pi)^{3/2}}\int\frac{d^3k}{\sqrt{2\omega_k}}\Big(a^+_\mu(\vec k)
e^{+ikx}+a^-_\mu(\vec k)e^{-ikx}\Big),
\end{equation}
where $k_0 = \omega_k>0$ is a (not necessarily specified) function of 
$|\vec k|$.
Inserting this into the expression for $H_0^*$ we find
\begin{eqnarray*}
&&\hspace{-1cm}
H_0^*=\frac12\int\frac{d^3k}{\sqrt{2\omega_k}}\int\frac{d^3q}{\sqrt{2\omega_q}}\\
&\times &
\Big(e^{it(\omega_k+\omega_q)-\frac i2\theta^{\mu\nu}k_\mu q_\nu}
\delta^3(\vec k+\vec q)\big(k_\mu q_\mu a^{\nu+}(\vec k)a^+_\nu(\vec q)\\
&&+(\frac{\alpha-1}{2\alpha})(
-k^\mu q_\nu a^+_\mu(\vec k)a^+_\nu(\vec q)
-k_\nu q^\mu a^+_\nu(\vec k)a^+_\mu(\vec q))\big)\\
&&+e^{-it(\omega_k+\omega_q)-\frac i2\theta^{\mu\nu}k_\mu q_\nu}
\delta^3(\vec k+\vec q)\big(k_\mu q_\mu a^{\nu-}(\vec k)a^-_\nu(\vec q) \\
&&+ (\frac{\alpha-1}{2\alpha})(
-k^\mu q_\nu a^-_\mu(\vec k)a^-_\nu(\vec q)
-k_\nu q^\mu a^-_\nu(\vec k)a^-_\mu(\vec q))
\big)\\
&&+e^{it(\omega_k-\omega_q)+\frac i2\theta^{\mu\nu}k_\mu q_\nu}
\delta^3(\vec k-\vec q)\big(-k_\mu q_\mu a^{\nu+}(\vec k) a_\nu^-(\vec q)
\\
&&+(\frac{\alpha-1}{2\alpha})(+k^\mu q_\nu a^+_\mu(\vec k)a^-_\nu(\vec q)
+k^\mu q_\nu a^-_\nu(\vec k)a^+_\mu(\vec q))\big)\\&&+
e^{-it(\omega_k-\omega_q)+\frac i2\theta^{\mu\nu}k_\mu q_\nu}
\delta^3(\vec k-\vec q)\big(-k_\mu q_\mu a^{\nu-}(\vec k) a_\nu^+(\vec q)
\\&&+(\frac{\alpha-1}{2\alpha})(+k^\mu q_\nu a^-_\mu(\vec k)a^+_\nu(\vec q)
+k^\mu q_\nu a^+_\nu(\vec k)a^-_\mu(\vec q))\big)\big)\Big).
\end{eqnarray*}
Using now the delta functions we find
\begin{eqnarray*}
&&\hspace{-1cm}
H_0^*=\frac12\int\frac{d^3k}{\sqrt{2\omega_k}}\int\frac{d^3q}{\sqrt{2\omega_q}}\\
&\times&
\Big(e^{it(\omega_k+\omega_q)-\frac i2\theta^{\mu\nu}k_\mu q_\nu}
\delta^3(\vec k+\vec q)\big(\frac12(k^2 a^{\nu+}(\vec k)a^+_\nu(\vec q)
+q^2 a_\nu^+(\vec k)a^{\nu+}(\vec q))\\
&&+(\frac{\alpha-1}{2\alpha})(
-k^\mu k^\nu a^+_\mu(\vec k)a^+_\nu(\vec q)
-q^\nu q^\mu a^+_\nu(\vec k)a^+_\mu(\vec q))\big)\\
&&+e^{-it(\omega_k+\omega_q)-\frac i2\theta^{\mu\nu}k_\mu q_\nu}
\delta^3(\vec k+\vec q)\big(\frac12(k^2 a^{\nu-}(\vec k)a^-_\nu(\vec q)
+q^2 a_\nu^-(\vec k)a^{\nu-}(\vec q) \\
&& + (\frac{\alpha-1}{2\alpha})(
-k^\mu k^\nu a^-_\mu(\vec k)a^-_\nu(\vec q)
-q^\nu q^\mu a^-_\nu(\vec k)a^-_\mu(\vec q))
\big)\\
&&+\delta^3(\vec k-\vec q)\big(-k_\mu k_\mu a^{\nu+}(\vec k) a_\nu^-(\vec k)
\\
&&+(\frac{\alpha-1}{2\alpha})(+k^\mu k_\nu a^+_\mu(\vec k)a^-_\nu(\vec k)
+k^\mu k_\nu a^-_\nu(\vec k)a^+_\mu(\vec k))\big)\\
&&+ \delta^3(\vec k-\vec q)\big(-k_\mu k_\mu a^{\nu-}(\vec k) a_\nu^+(\vec k)
\\&&+(\frac{\alpha-1}{2\alpha})(+k^\mu k_\nu a^-_\mu(\vec k)a^+_\nu(\vec k)
+k^\mu k_\nu a^+_\nu(\vec k)a^-_\mu(\vec k))\big)\big)\Big).
\end{eqnarray*}
With the equation of motion (\ref{eomgf}) expressed in terms of $a_\mu^{\pm}$,
\begin{equation}
k^2a_\mu^\pm(\vec k)-(\frac{\alpha-1}{\alpha})k_\mu(k^\nu a_\nu^\pm(\vec k))=0,
\end{equation}
the terms with the non zero exponentials vanish. The remaining terms 
simplify considerably with the help of the equation of motion. So we get
\begin{eqnarray}
H_0^* &=& \frac12\int \frac{d^3k}{2\omega_k}\big(-k_\mu k_\mu a^{\nu+}(\vec k)
a^-_{\nu}(\vec k) + k^2 a^+_\nu(\vec k)a^-_{\nu}(\vec k)
\nonumber\\&&\qquad\qquad-k_\mu k_\mu a^{\nu-}(\vec k)
a^+_{\nu}(\vec k) + k^2 a^-_\nu(\vec k)a^+_{\nu}(\vec k)\big)\nonumber\\
&=&\int \frac{d^3k}{2\omega_k}\big(-\vec k^2(a_0^+a_0^-+a_0^-a_0^+)
+\omega_k^2(a_i^+a_i^-+a_i^-a_i^+)\big) = H_0.
\end{eqnarray}
Quantisation can now be performed in the usual way by imposing appropriate 
commutator relations (e.g. for $\alpha=1$, Feynman gauge)
\begin{equation}
[a^-_\rho(\vec k), a^+_\mu (\vec k\,')] = -g_{\rho\mu}\delta^{3}(\vec k
-\vec k\,')\ .
\end{equation}

\subsection{BRST-Symmetry}

The free part of the BRST-expanded Lagrangian on a noncommutative space
reads
\begin{eqnarray}
L_0 =\int d^3 x \Big(-\frac14 f_{\mu\nu}f^{\mu\nu}+
B(\partial^\mu A_\mu+
\frac\alpha2 B) + \partial^\mu \bar c\partial_\mu c\Big).
\end{eqnarray} 
The equations of motion read
\begin{eqnarray}
\frac{\partial{\mathcal L}_0}{\partial A^\mu}&=&\square A_\mu-\partial_\mu(\partial^\nu A_\nu)
-\partial_\mu B = 0, \nonumber\\
\frac{\partial{\mathcal L}_0}{\partial B}&=& \partial^\mu A_\mu +\alpha B = 0.
\end{eqnarray}
We postpone the treatment of the ghost sector, which in the free theory
decouples from the gauge field sector anyway.
In order to construct the Hamiltonian we have
\begin{eqnarray} 
&&\frac{\partial {\mathcal L}_0}
{\partial \dot A^i} =: \Pi_i = f_{i0},\nonumber\\
&&\frac{\partial{\mathcal L}_0}{\partial \dot A^0} =: \Pi_0= B,
\qquad \frac{\partial {\mathcal L}_0}{\partial \dot B} =: \Pi_B = 0.
\end{eqnarray}
The latter two equations are primary constraints. Since their Poisson
bracket is not weakly zero,
\begin{equation}
\label{constraints}
\{\phi_1(\vec x),\phi_2(\vec x')\}_{PB}=
\{\Pi_0(\vec x)-B(\vec x),\Pi_B(\vec x')\}_{PB}=
-\delta^3(\vec x - \vec x'),
\end{equation}
they are second class constraints. 

Now, in order to write down 
the corresponding \emph{noncommutative} Hamiltonian,
we firstly define the symmetrized $\star_s$-product,
\begin{equation}
A\star_s B = \frac12(A\star B \pm B\star A),
\end{equation}
where the sign is positive for usual fields and negative for 
Grassmann valued fields. Again, the 
use of this $\star_s$-product is crucial.

The total noncommutative Hamiltonian \cite{Dirac}
thus 
reads (with use of $\dot A_i\star_s\Pi^i = 
(\partial_i A_0-\Pi_i)\star_s\Pi^i$ and partial integration)
\begin{eqnarray}
H_T^* &=& \int d^3x\Big( \dot A_\mu\star_s\Pi^\mu +\dot B\star_s\Pi_B 
-{\mathcal L}_0 +\lambda'_1\star_s\phi_1+\lambda'_2\star_s\phi_2\Big)
\nonumber\\
&=& \int d^3x \Big(\underbrace{(\lambda'_1+\dot A_0)}_
{\lambda_1}\star_s(\Pi^0-B)+\underbrace{(\lambda'_2+\dot B)}_{\lambda_2}
\star_s\Pi_B -A_0\star_s\partial_i\Pi^i\nonumber\\&&\qquad
-B\star_s\partial_i A^i -\frac\alpha2 B\star B +\frac12\Pi^i\star \Pi^i
+\frac14 f^{ij}\star f^{ij}\Big),
\end{eqnarray}
where $\phi_1$ and $\phi_2$ denote the constraints defined in
(\ref{constraints}), $\lambda'_1$ and $\lambda'_2$ are Lagrange multiplier.
Since the constraints should be preserved in time, we find conditions 
for $\lambda_i$,
\begin{eqnarray}
\{H_T^*,\phi_1\}_{PB} &=& \lambda_2-\partial_i\Pi^i = 0,\nonumber \\{} 
\{H_T^*,\phi_2\}_{PB} &=& -\lambda_1 -\partial_i A^i-\alpha B =0.
\end{eqnarray}
According to Dirac \cite{Dirac}, for quantisation the second 
class constraints are imposed as strong operator equations. This is only
possible after elimination of the unphysical degrees of freedom 
corresponding to the second class constraints. Clearly, these degrees of 
freedom are simply $B,\ \Pi_B$.

So, with $\Pi_B = 0$ and $B=\Pi^0$
we get the quantisable noncommutative Hamiltonian
\begin{eqnarray}
&&{H'}^* = \int d^3x\Big( -A_0\star_s\partial_i\Pi^i
-\Pi^0\star_s\partial_i A^i\nonumber\\&&\qquad\qquad -\frac\alpha2 \Pi^0\star \Pi^0
 +\frac12\Pi^i\star \Pi^i
+\frac14 f^{ij}\star f^{ij}\Big).
\end{eqnarray}
With use of the Hamiltonian equations of motion for the fields,
\begin{eqnarray}
&&\dot A_0 = \frac{\delta {H'}^*}{\delta \Pi^0}=
-\partial_i A^i -\alpha \Pi^0, \qquad
\dot A_i = \frac{\delta {H'}^*}{\delta \Pi^i}= \partial_i A^0 -\Pi_i,
\end{eqnarray}
we may express the field momenta by the fields and their time derivative. 
Inserting this yields exactly the Hamiltonian
(\ref{Hamiltongf}) we have found for the gauge fixed theory:
\begin{eqnarray}
&& {H'}^* = \int d^3x\Big((-A^0\star_s\partial^i\dot A^i 
+A^0\star_s\partial^i\partial^iA^0)\nonumber\\&&
\qquad +(\frac1\alpha\partial_i A^i\star
\partial_j A^j+\frac1\alpha\dot A^0\star_s\partial_j A^j)\nonumber\\&&\qquad+
(-\frac1{2\alpha}\partial_i A^i\star\partial_j A^j-\frac1{2\alpha}
\dot A^0\star\dot A^0 - \frac2{2\alpha}\partial_i A^i\star_s\dot A^0)
\nonumber\\&&\qquad
+(\frac12\dot A^i\star\dot A^i +\frac12\partial^i A^0\star\partial^i A^0
-\dot A^i\star_s\partial^i A^0)\nonumber\\&&\qquad
 +(\frac12\partial^i A^j\star\partial^i A^j
-\frac12\partial^i A^j\star\partial^j A^i)\Big)\nonumber\\
&&=\frac12\int d^3x \Big(\dot A_i\star \dot A_i -\dot A_0\star \dot A_0
+\partial_j A_i \star\partial_j A_i-\partial_j A_0 \star\partial_j A_0
\nonumber\\&&\qquad
+(\frac{\alpha-1}{\alpha})(\dot A_0\star \dot A_0
-(\partial_i A_i)\star(\partial_j A_j))\Big).
\end{eqnarray}
Note: The elimination of the $B$-field does not spoil our considerations 
with respect to the construction of perturbation theory, since 
the $B$ field has no interaction vertex. 

Now for $c,\ \bar c$ the situation is very simple,
\begin{eqnarray}
L_{\phi\Pi} = \int d^3x {\mathcal L}_{\phi\Pi} =
\int d^3x \partial^\mu\bar c \star_s \partial_\mu c.
\end{eqnarray}
The equations of motion and the momenta are
\begin{eqnarray}
\frac{\partial{\mathcal L} _{\phi\Pi}}{\partial \bar c} = -\square c =0, \qquad
\frac{\partial{\mathcal L}_{\phi\Pi}}{\partial c} = \ \square \bar c =0,\nonumber \\
\frac{\partial{\mathcal L}_{\phi\Pi}}{\partial\dot{\bar c}}=\Pi_{\bar c}=\dot c,\qquad
\frac{\partial{\mathcal L}_{\phi\Pi}}{\partial\dot c} = \Pi_c = -\dot{\bar c}.
\end{eqnarray}
There are no constraints. For the noncommutative Hamiltonian we have
\begin{eqnarray}
H_{\phi\Pi}^* &=& \int d^3x (\dot{\bar c}\star_s\Pi_{\bar c}+
\dot c\star_s \Pi_c +\Pi_c\star_s\Pi_{\bar c}+
\partial_i\bar c\star_s\partial_i c) \nonumber\\
&=& \int d^3x (\dot{\bar c}\star_s\dot c - \dot c \star_s\dot{\bar c}
-\dot{\bar c}\star_s\dot c +\partial_i\bar c\star_s\partial_i c)
\nonumber\\&=& \int d^3x(\dot{\bar c}\star_s\dot c +\partial_i
\bar c\star_s\partial_i c).
\end{eqnarray}
We check time independence of $H_{\phi\Pi}^*$,
\begin{eqnarray}
\dot H_{\phi\Pi}^*= \int d^3x 
\Big((\ddot{\bar c}-\partial_i\partial_i \bar c)\star_s\dot c
+\dot{\bar c}\star_s(\ddot c - \partial_i\partial_i c)\Big) = 0.
\end{eqnarray}
Using the following ansatz for $c, \bar c$,
\begin{eqnarray}
c(x)&=&\frac1{(2\pi)^{3/2}}\int\frac{d^3k}{\sqrt{2\omega_k}}\Big(c^+(\vec k)
e^{ikx}+c^-(\vec k)e^{-ikx}\Big),\nonumber\\
\bar c(x)&=&\frac1{(2\pi)^{3/2}}\int\frac{d^3k}{\sqrt{2\omega_k}}\
\Big(\bar c^+(\vec k)e^{ikx}-\bar c^-(\vec k)e^{-ikx}\Big)
\end{eqnarray} 
(note that $\bar c(x)$ is here imaginary) 
and the Poisson bracket for Grassmann fields, $\{c(\vec x),\Pi_c(\vec x')\}=\{\bar c(\vec x),\Pi_{\bar c}
(\vec x')\}=-\delta^3(\vec x-\vec x')$, we find
\begin{equation}
\{\bar c^+(\vec k), c^-(\vec k'\} = 
\{\bar c^-(\vec k), c^+(\vec k')\}=
-i\delta^3(\vec k-\vec k').
\end{equation}
For $H_{\phi\Pi}^*$ we find with the help of the equations of motion
(with $k_0=\omega_k=\sqrt{\vec k^2}$, so that $k_\mu k_\mu = 2\omega_k^2$)
\begin{eqnarray}
&&H_{\phi\Pi}^* = \int d^3x \partial_\mu{\bar c}(x)\star_s\partial_\mu c(x) 
\nonumber\\
&&= \int \frac{d^3x}{(2\pi)^3}\int\frac{d^3k\,d^3k'}
{2\sqrt{\omega_k\omega_{k'}}}\Big((+ik_\mu\bar c^+(\vec k)e^{ikx}
+ik_\mu\bar c^-(\vec k)e^{-ikx})
\nonumber\\&&\qquad\qquad\star_s(+ik'_\mu c^+(\vec k')e^{ik'x}
-ik'_\mu c^-(\vec k')e^{-ik'x})\Big)\nonumber\\&&=
\int d^3k\omega_k\Big(\bar c^+(\vec k)c^-(\vec k)+
c^+(\vec k)\bar c^-(\vec k)\Big) = H_{\phi\Pi}.
\end{eqnarray}
We find that noncommutativity does not spoil the free theory.
Quantisation 
is done by the replacement of the Poisson brackets by commutators,   
\begin{equation}
\{\ ,\ \}_{PB} \Rightarrow -i[\ ,\ ]\ ,
\end{equation}
which again leads to the well known commutator relations between
annihilation and creation operators of fields.

\section{Conclusion}

We have constructed perturbation theory on a noncommutative space
from the beginning. We have discussed a general non-local deformation of the 
free Hamiltonian. For some of the considered deformations, such as the canonical one
\cite{Doplicher:1995tu}, the free Hamiltonian is unaffected by the deformation. 
In the canonical case, this is even true for gauge field theory as shown in
the previous section.
This implies that IPTOPT as developed in 
\cite{Doplicher:1995tu,Liao:2002xc,Denk:2003jj,Bozkaya:2002at} is consistent. However, 
the application of these techniques to Gaussian non-localities 
\cite{Bahns:2003vb,Denk:2004pk} seems to be somewhat artificial since 
here we could show that the introduction of the non-localities into the 
free Hamiltonian would alter it significantly when assuming the usual free 
field equations. Thus, either one  has to leave the free Hamiltonian untouched  
\cite{Denk:2004pk} or one has to develop an 
appropriate free theory.

Furthermore, the discrepancies  between IPTOPT and the NPIA present for 
non-localities in time have been discussed in some detail. The main reason is the problem 
of passing from non-local products  given in the Dirac  picture to the Heisenberg picture.  
One has to alter these products in order for them to be consistent. But usually one compares 
situations where one deals with the same product in both pictures, which cannot give 
the same as soon as non-localities in time are involved. We also want to point out that 
the use of $L_I$ instead of $H_I$ in combination with IPTOPT is not justified 
by path integrals as soon as non-localities involve time.

The main motivation for studying IPTOPT is its unitarity \cite{Liao:2002pj} 
and the well behaving UV/IR mixing \cite{Fischer:2003jh}. The disadventages 
are the violation of causality and the higher complexity of the Feynman rules. 
The situation is vice versa for the NPIA, where the implied time ordering 
respects causality.
The violation of unitarity in NPIA is a severe problem, 
whereas it seems to be absent in IPTOPT.

\subsection*{Acknowledgement}

We are grateful to Manfred Schweda for intensive discussions and many very
valuable remarks.
This work has been supported by DOC (predoc program of the \"Osterreichische
Akademie der Wissenschaften) (S.D.) and by "Fonds zur F\"orderung der
Wissenschaftlichen Forschung" (FWF), under contract P15015-N08 (V.P.) and
P15463-N08 (M.W.).

\bibliographystyle{../latex-styles/JHEP-3}

\providecommand{\href}[2]{#2}
\begingroup\raggedright


\end{document}